\documentclass[a4paper,10pt]{article}
\usepackage[pdftex]{graphicx,hyperref}
\usepackage{amsmath,amsfonts,amssymb}
\usepackage{wasysym}
\usepackage{fullpage}
\usepackage{enumerate}
\usepackage{sidecap}
\usepackage{lscape}
\usepackage{authblk}
\usepackage{appendix}
\usepackage{setspace}
\title{\bf{Socioeconomic correlations and stratification in social-communication networks}}

\author[a]{Yannick Leo}
\author[a]{Eric Fleury}
\author[b,c]{J. Ignacio Alvarez-Hamelin}
\author[d]{Carlos Sarraute}
\author[a,*]{M\'arton Karsai}

\affil[a]{\small{Univ Lyon, ENS de Lyon, INRIA, CNRS, UMR 5668, IXXI, 69364 Lyon, France}}
\affil[b]{\small{Univerisidad de Buenos Aires, Facultad de Ingenier\'ia, Av. Paseo Col\'on 850, C1063ACV Buenos Aires, Argentina.}}
\affil[c]{\small{CONCIET, Buenos Aires, Argentina.}}
\affil[d]{\small{Grandata Labs, Bartolome Cruz 1818 Vicente Lopez. Buenos Aires, Argentina}}
\affil[*]{\small{marton.karsai@ens-lyon.fr}}

\date{}

\begin{document}

\maketitle

\begin{abstract}
The uneven distribution of wealth and individual economic capacities are among the main forces which shape modern societies and arguably bias the emerging social structures. However, the study of correlations between the social network and economic status of individuals is difficult due to the lack of large-scale multimodal data disclosing both the social ties and economic indicators of the same population. Here, we close this gap through the analysis of coupled datasets recording the mobile phone communications and bank transaction history of one million anonymised individuals living in a Latin American country. We show that wealth and debt are unevenly distributed among people in agreement with the Pareto principle; the observed social structure is strongly stratified, with people being better connected to others of their own socioeconomic class rather than to others of different classes; the social network appears with assortative socioeconomic correlations and tightly connected ``rich clubs''; and that egos from the same class live closer to each other but commute further if they are wealthier. These results are based on a representative, society-large population, and empirically demonstrate some long-lasting hypotheses on socioeconomic correlations which potentially lay behind social segregation, and induce differences in human mobility.
\end{abstract}

Socioeconomic imbalances, which universally characterise all modern societies~\cite{Piketti2014Capital, Sernau2012Social}, are partially induced by the uneven distribution of economic power between individuals. Such disparities are among the key forces behind the emergence of social inequalities~\cite{Sernau2012Social,Hurst2015Social}, which in turn leads to social stratification and spatial segregation in social structures characterised by correlations between the social network, living environment, and socioeconomic status of people. Although this hypothesis has been drawn long time ago~\cite{Grusky2011Theories}, the empirical observation of spatial, socioeconomic, and structural correlations in large social systems has been difficult as it requires simultaneous access to multimodal characters for a large number of individuals. Our aim in this study is to find evidence of social stratification through the analysis of a combined large-scale anonymised dataset that disclose simultaneously the social interactions, frequent locations, and the economic status of millions of individuals.

The identification of socioeconomic classes is among the historical questions in the social sciences with several competing hypothesis proposed on their structure and dynamics~\cite{Giddens1973Class}. One broadly-accepted definition identifies lower, middle, and upper classes~\cite{Akhbar2010Class,Brown2009Social,Stark2007Sociology, Gilbert2002The, Stiglitz2012The} based on the socioeconomic status of individuals. These classes can be further used to indicate correlations characterising the social system. People who live in the same neighbourhood may belong to the same class, and may have similar levels of education, jobs, income, ethnic background, and may even share common political views. These similarities together with homophily, i.e. the tendency that people build social ties with similar others \cite{McPherson2001Birds,Lazarsfeld1954Friendship}, strongly influence the structure of social interactions and have indisputable consequences on the global social network as well. The coexistence of social classes and homophily may lead to a strongly stratified social structure where people of the same social class tend to be better connected among each other, while connections between different classes are less frequent than one would expect from structural characteristics only~\cite{Grusky2011Theories,Doob2015Social,Saunders1990Social}. These correlations may further determine the living environment and mobility of people leading to spatial segregation and specific commuting patterns characterising people from similar social classes \cite{Carra2016Modelling,Sim2015Great,Iceland2006Does}.

The observation of such correlations should be possible through the analysis of the social structure~\cite{Wasserman1994Social}. Research on social networks has recently been accelerated through the advent of new technologies which allow the collection of detailed digital footprints of interactions of large number of people~\cite{Lohr2012The,lazerscience2009}. These advancements lead us to observations on the heterogeneous, structurally, spatially, and temporally correlated social networks~\cite{Abraham2010Computational,Newman2003The}, and to the identification of social mechanisms driving their evolution~\cite{Hedstrom1998Social,Holme2015Mechanistic}. However, although such datasets may contain some information about individual characteristics, they commonly miss one important dimension: they do not provide any direct estimator of the economic status of people, which could strongly influence their connection preferences and may determine the social position of an individual in the global social network. Coarse-grained information about people's economic status are typically provided as statistical census measures without disclosing the underlying social structure, or by social surveys \cite{Campbell1986Social} covering a small and less representative population.

In this paper, we aim to close this gap through the analysis of a combined dataset collecting the social interactions, proxy location, and economic situation of a large set of individuals. More precisely, we  analyse the transaction and purchase history coupled with time-resolved, spatially detailed mobile phone interactions of millions of anonymised inhabitants of a Latin American country over 8 months (for a detailed data description see Data and Materials). After introducing precise indicators of economic status, we show that not only individual income but also debt is distributed unevenly in accordance with the Pareto principle. Through the detection of homophilic correlations in the social structure, we provide strong empirical evidence of the stratified intra- and inter-class structure of the social network, and the existence of assortative socioeconomic correlations and ``rich clubs''. Finally we present quantitative results about the relative spatial distribution and typical commuting distances of people from different socioeconomic classes.

\section*{Results}

The full description of one's socioeconomic status is rather difficult as it is characterised not only by quantitative features but also related to one's social or cultural capital \cite{Bourdieu1984}, reputation, or professional skills. However, we can estimate socioeconomic status by assuming a correlation between one's social position and economic status, which can be approximated by following the network position and financial development of people. This approach in turn not only gives us a measure of an individual's socioeconomic status but can also help us to draw conclusions about the overall distribution of socioeconomic potential in the larger society.

\subsection*{Economic status indicators}

Our estimation of an individual's economic status is based on the measurement of consumption power. We use a dataset which contains the amount and type of daily debit/credit card purchases, monthly loan amounts, and some personal attributes such as age, gender, and zip code of billing address of $\sim 6$ million anonymised customers of a bank in the studied country over $8$ months (for further details see Data and Materials). In addition, for a smaller subset of clients, the data provide the precise salary and total monthly income that we use for verification purposes as explained later.

\begin{figure}[t!]
\centering
\includegraphics[width=0.94\textwidth,angle=0]{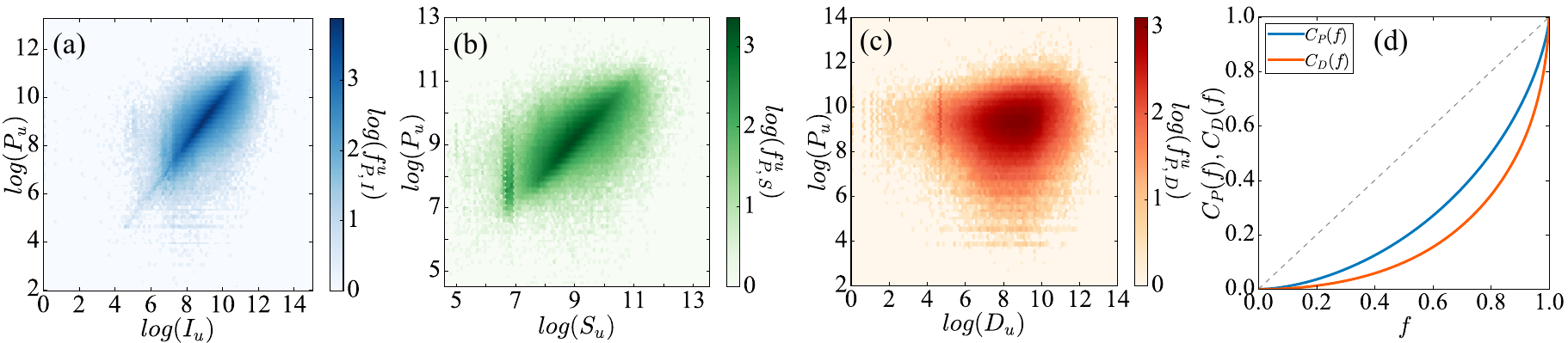}
\caption{Correlations and distributions of individual economic indicators. The heat maps show correlations between the average monthly purchase $P_u$ and (a) average income $I_u$, (b) average salary $S_u$, and (c) average monthly debt $D_u$ for (a) 625,412 (b) 389,567 and (c) 339,288 customers who have (accordingly) both corresponding measures available. Colours in panels (a-c) depicts the logarithm of the fraction of customers with the given measures. (d) Cumulative distributions of $P_u$ (blue line) and $D_u$ (orange line) as functions of sorted fraction $f$ of individuals. Distributions were measured for $6,002,192$ (resp. $339,288$) individuals from whom AMP (resp. AMD) values were available. Dashed line shows the case of the perfectly balanced distribution.
\label{fig:1}}
\end{figure}

By following the purchase history of each individual, we estimate their economic position from their average amount of debit card purchases. More precisely, for an individual $u$ who spent a total amount of $p_u(t)$ in month $t$, we estimate his/her average monthly purchase (AMP) as
\begin{equation}
P_u=\frac{\sum_{t\in T} p_u(t)}{|T|_u},
\end{equation}
where $|T|_u$ corresponds to the number of active months of the user (with at least one purchase). In order to verify this individual economic indicator we check its correlations with other indicators, such as the salary $S_u$ (defined as the average monthly salary of individual $u$ over the observation period $T$) and the income $I_u$ (defined as the average total monthly income including salary and other incoming bank transfers). We find strong correlations between individual AMP $P_u$ and income $I_u$ with a Pearson correlation coefficient $r\approx 0.758$ ($p<.001$, $SE=7.33\times 10^{-4}$) (for correlation heat map see Fig.\ref{fig:1}a), and also between $P_u$ and salary $S_u$ with  $r\approx 0.691$  ($p<.001$, $SE=9.695\times 10^{-4}$) (see Fig.\ref{fig:1}b). Note that direct economic indicators, such as $I_u$ and $S_u$, are available only for a smaller subset of users (for exact numbers see Fig.\ref{fig:1} caption), thus for the present study we decided to use $P_u$ since this measure is available for the whole set of users.

At the same time we are interested in an equivalent indicator which estimates the financial commitments of individuals. We define the average monthly debt (AMD) of an individual $u$ by measuring
\begin{equation}
D_u=\frac{\sum_{t\in T} d_u(t)}{|T|_u},
\end{equation}
where $d_u(t)$ indicates the debt of individual $u$ in month $t\in T$ and $|T|_u$ is the number of active months where the user had debt. Arguably individual debt could depend on the average income and thus on the AMP of a person due to the loaning policy of the bank. Interestingly, as demonstrated in Fig.\ref{fig:1}c, we found weak correlations between AMP and AMD with a small coefficient $r \approx 0.104$ ($p<.001$, $SE=2.48\times 10^{-3}$), which suggests that it is worth to treat these two indicators independently.

\subsection*{Overall socioeconomic imbalances}

The distribution of an individual economic indicator may disclose signs of socioeconomic imbalances on the population level. This hypothesis was first suggested by V. Pareto and later became widely known as the law named after him~\cite{Pareto1971Manual}. The present data provide a straightforward way to verify this hypothesis through the distribution of individual AMP. We measured the normalised cumulative function of AMP for $f$ fraction of people sorted by $P_u$ in an increasing order:
\begin{equation}
C_P(f)=\frac{1}{\sum_u P_u}\sum_{f} P_u
\end{equation}
We computed this distribution for the $6,002,192$ individuals assigned with AMP values. This function shows (see Fig.\ref{fig:1}d blue line) that AMP is distributed with a large variance, i.e., indicating large economical imbalances just as suggested by the Pareto's law. A conventional way to quantify the variation of this distribution is provided by the Gini coefficient $G$ \cite{Gastwirth1972The}, which characterises the deviation of the $C_P(f)$ function from a perfectly balanced situation, where wealth is evenly distributed among all individuals (diagonal dashed line in Fig.\ref{fig:1}d). In our case we found $G_P\approx 0.461$, which is relatively close to the World Bank reported value $G=0.481$ for the studied country~\cite{World2010Gini}, and corresponds to a Pareto index~\cite{Souma2000Physics} $\alpha=1.315$. This observation indicates a $0.73:0.27$ ratio characterising the uneven distribution of wealth, i.e., that the $27\%$ of people are responsible for the $73\%$ of total monthly purchases in the observed population. Note that these values are close to the values $G=0.6$ and $80:20$, which were suggested by Pareto. 

At the same time we have characterised the distribution of individual AMD by measuring the corresponding $C_D(f)$ function as shown in Fig.\ref{fig:1}d (orange line) for $339,288$ individuals for whom AMD values were available. It indicates even larger imbalances in case of debt with a Gini coefficient $G_D \approx 0.627$ and $\alpha=1.140$ indicating $19\%$ of the population to be actually responsible for $81\%$ of overall debt in the country. This observation suggests that Pareto's hypothesis holds not only for the distribution of purchases but for debt as well. Note that similar distribution of debt of bankrupt companies has been reported \cite{AoyamaPareto2000}.

\subsection*{Class definition and demographic characters}

The economic capacity of individuals arguably correlates with their professional occupation, education level, and housing, which in turn determine their social status and environment. At the same time status homophily~\cite{McPherson2001Birds,Lazarsfeld1954Friendship}, i.e., people's tendency to associate with others of similar social status, has been argued to be an important mechanism that drives the creation of social ties. Our hypothesis is that these two effects, diverse socioeconomic status and status homophily, potentially lead to the emergence of a stratified structure in the social network where people of the same social class tend to be better connected among themselves than with people from other classes. A similar hypothesis has been suggested earlier~\cite{Bottero2005Stratification} but its empirical verification has been impossible until now as this would require detailed knowledge about the social structure and precise estimators of individual economic status. In the following, our main contribution is to clearly identify signatures of social stratification in a representative society-level dataset, that contains information on both the social network structure and the economic status of people.

In order to investigate signatures of social stratification, we combine the bank transaction data with data disclosing the social connections between the bank's customers. To identify social ties, we use a mobile communication dataset, provided by one mobile phone operator in the country, with a customer set that partially overlaps with the user set found in the bank data (for details on data matching policy see Data and Materials). To best estimate the social network, we connect people who at least once communicated with each other via call or SMS during the observation period of 21 months between January 2014 and September 2015, but we remove non-human actors, such as call centres and commercial communicators, by using a recursive filtering method. For the purpose of our study we select all mobile phone users who appear as customers in the bank dataset and take the largest connected component of the intersection graph. After this procedure we obtain a social network with $|E|=1,960,239$ links and $N=992,538$ nodes, each corresponding to an individual with a valid non-zero AMP value $P_u$. For further details about the datasets, their combinations, filtering, and network construction see Data and Materials. 

\begin{figure}[t!]
\centering
\includegraphics[width=0.94\textwidth,angle=0]{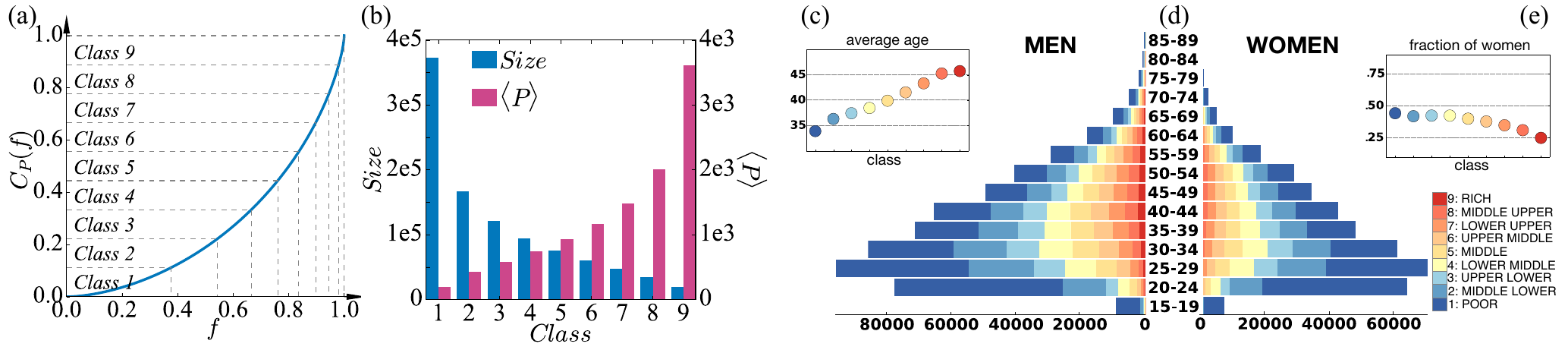}
\caption{Social class characteristics (a) Schematic demonstration of user partitions into 9 socioeconomic classes by using the cumulative average monthly purchase (AMP) function $C_P(f)$. Fraction of individuals belonging to a given class ($x$ axis) have the same sum of AMP $(\sum_u P_u)/n$ ($y$ axis) for each class. (b) Number of egos (blue), and the average AMP $\langle P \rangle$ (in USD \cite{Currency}) per individual (pink) in different classes. (c) Average age of different classes. (d) Age pyramids for men and women with colours indicating the corresponding socioeconomic groups and with bars proportional to absolute numbers. (e) Fraction of women in different classes.
\label{fig:2}}
\end{figure}

Taking each individual in the selected social network, we assign each of them into one of $n=9$ socioeconomic classes based on their individual AMP values. This classification is defined by sorting individuals by their AMP, take the cumulative function $C_P(f)$ of AMP, and cut it in $n$ segments such that the sum of AMP in each class is equal to $(\sum_u P_u)/n$ (as shown in Fig.\ref{fig:2}a). Our selection of nine distinct classes is based on the common three-stratum model \cite{Brown2009Social,Akhbar2010Class}, which identifies three main social classes (lower, middle, and upper), and  three sub-classes for each of them~\cite{Saunders1990Social}. More importantly, this way of classification relies merely on individual economic estimators, $P_u$, and naturally partition individuals into classes with decreasing sizes, and increasing $\langle P \rangle$ per capita average AMP values for richer groups (for exact values see Fig.\ref{fig:2}b)\cite{Currency}. To explore the demographic structure of the classes we used data on the age and gender of customers. We draw the population pyramids for men and women in Fig.\ref{fig:2}d with colour-bars indicating the number of people in a given social class at a given age. We found a positive correlation between social class and average age, suggesting that people in higher classes are also older on average (see Fig.\ref{fig:2}c). In addition, our data verifies the presence of gender imbalance as the fraction of women varies from $0.45$ to $0.25$ by going from lower to upper socioeconomic classes (see Fig.\ref{fig:2}e).

\subsection*{Structural correlations and social stratification}

Using the above-defined socioeconomic classes and the social network structure, we turn to look for correlations in the inter-connected class structure. To highlight structural correlations, such as the probability of connectedness, we use a randomised reference system. It is defined as the corresponding configuration network model structure where we take the original social network, select random pairs of links and swap them without allowing multiple links and self loops. In order to remove any residual correlations we repeated this procedure $5\times |E|$ times. This randomisation keeps the number of links, individual economic indicators $P_u$, and the assigned class of people unchanged, but destroys any structural correlations in the social structure and consequently between socioeconomic layers as well. In each case, we repeat this procedure for $100$ times and present results averaged over the independent random realisations. Taking the original (resp. randomised) network we count the number of links $|E(s_i,s_j)|$ (resp. $|E_{rn}(s_i,s_j)|$) connecting people in different classes $s_i$ and $s_j$.  After repeating this procedure for each pair of classes in both networks, we take the fraction:
\begin{equation}
L(s_i,s_j)=\frac{|E(s_i,s_j)|}{|E_{rn}(s_i,s_j)|},
\label{eq:Lsisj}
\end{equation}
which gives us how many times more (or less) links are present between classes in the original structure as compared to the randomised one. Note that in the randomised structure the probability that two people from given classes are connected depends only on the number of social ties of the individuals and the size of the corresponding classes, but is independent of the effect of potential structural correlations. This way the comparison of the original and random structures highlights only the effect of structural correlations induced by status homophily or other tie creation mechanisms such as cyclic or triadic closure \cite{Kumpula2007Emergence}.

\begin{figure}[t!]
\centering
\includegraphics[width=0.94\textwidth,angle=0]{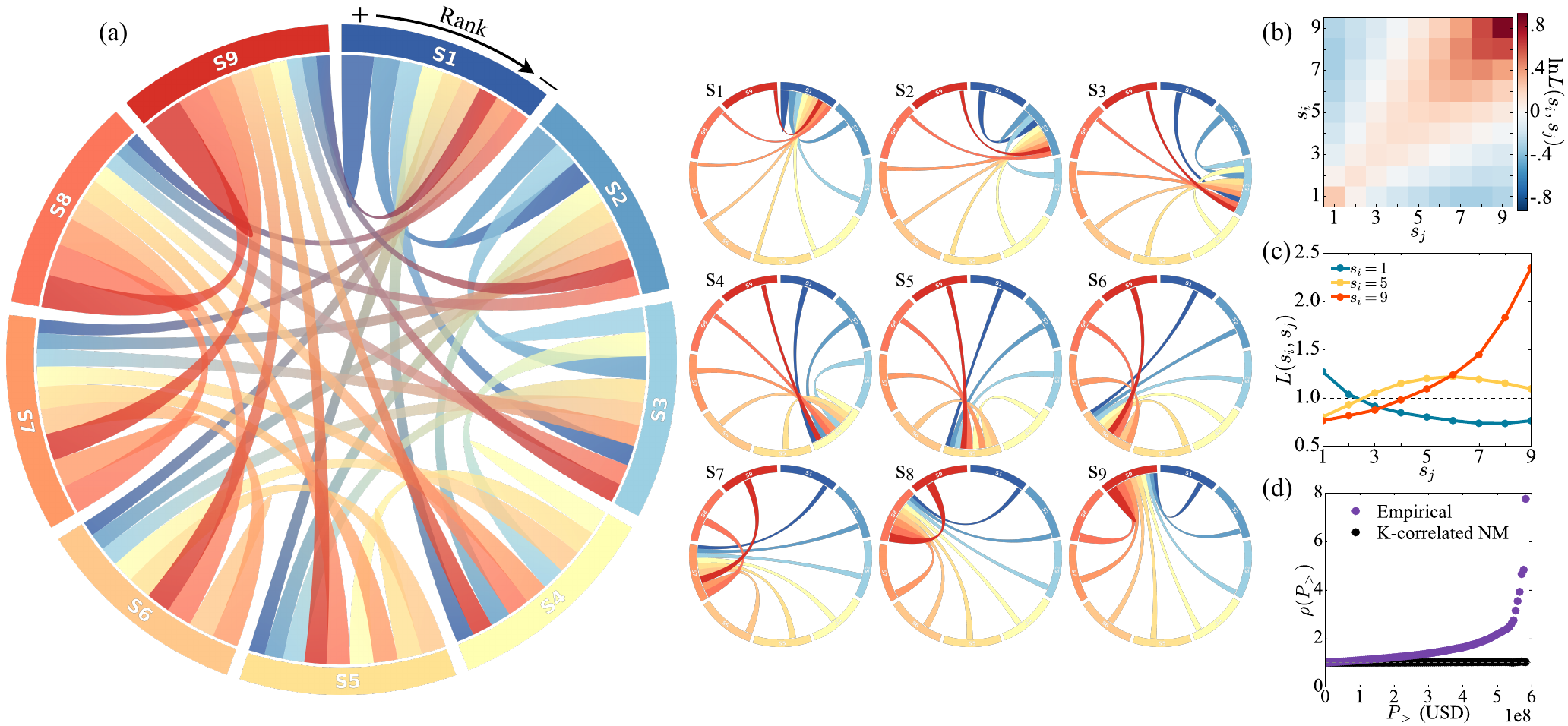}
\caption{Structural correlations in the socioeconomic network (a) Chord diagram of connectedness of socioeconomic classes $s_i$, where each segment represents a social class $s_i$ connected by chords with width proportional to the corresponding inter-class link fraction $\tilde{L}_{s_i}(s_j)$, and using gradient colours matched with opposite ends $s_j$. Note that the $\tilde{L}_{s_i}(s_j)=L(s_i,s_j)/\Sigma_{s_j}L(s_i,s_j)$ normalised fraction of $L(s_i,s_j)$ (in Eq.\ref{eq:Lsisj}) was introduced here to assign equal segments for each class for better visualisation. Chords for each class are sorted in decreasing width order in the direction shown above the main panel. On the minor chord diagrams of panel (a), graphs corresponding to each class are shown with non-gradient link colours matching the opposite end other than the selected class. (b) Matrix representation of $L(s_i,s_j)$ (for definition see Eq.\ref{eq:Lsisj}) with logarithmic colour scale. (c) The $L(s_i,s_j)$ function extracted for three selected classes ($1$ (blue), $5$ (yellow), and $9$ (red)). Panels (a)-(d) provide quantitative evidence on the stratified structure of the social network and the upward-biased connections of middle classes. (d) ``Rich-club'' coefficient $\rho(P_>)$ (definition see Eq.\ref{eq:RCC}) based on the empirical (purple), and a degree-correlated null model (black) networks. On the individual level the richest people of the population appear to be eight times more densely connected than expected randomly.
\label{fig:3}}
\end{figure}

From the chord diagram visualisation of this measure in Fig.\ref{fig:3}a, we can draw several conclusions. Note that for better visual presentation in Fig.\ref{fig:3}a we have normalised $L(s_i,s_j)$ and thus chord width indicates relative values $\widetilde{L}_{s_i}(s_j)=L(s_i,s_j)/\sum_{s_j}L(s_i,s_j)$ as compared to the origin class $s_j$ (as also explained in the figure caption). First, after sorting the chords of a given class $s_i$ in a decreasing $L(s_i,s_j)$ order, chords connecting a class to itself (self-links) appear always at top (or top 2nd) positions of the ranks. At the same time other top positions are always occupied by chords connecting to neighbouring social classes. These two observations (better visible in Fig.\ref{fig:3}a insets) indicate strong effects of status homophily and the existence of stratified social structure where people from a given class are the most connected with similar others from their own or from neighbouring classes, while connections with individuals from remote classes are least frequent. A second conclusion can be drawn by looking at the sorting of links in the middle and lower upper classes ($S4-S8$). As demonstrated in the inset of Fig.\ref{fig:3}a, people prefer to connect upward and tend to hold social ties with others from higher social classes rather than with people from lower classes.

These conclusions can be further verified by looking at other representations of the same measure. First we show a heat map matrix representation of Eq.\ref{eq:Lsisj} (see Fig.\ref{fig:3}b), where $L(s_i,s_j)$ values are shown with logarithmic colour scales. This matrix has a strong diagonal component verifying that people of a given class are always better connected among themselves (red) and with others from neighbouring groups, while social ties with people from remote classes are largely underrepresented (blue) as compared to the expected value provided by the random reference model. This again indicates the presence of homophily and the stratified structure of the socioeconomic network. The upward-biased inter-class connectivity can also be concluded here from the increase of the red area around the diagonal by going towards richer classes. These conclusions are even more straightforward from Fig.\ref{fig:3}c where the $L(s_i,s_j)$ is shown for three selected classes ($1$-poor, $5$-middle, and $9$-rich). These curves clearly indicate the connection preferences of the selected classes. Moreover, they show that richest people appear with the strongest homophilic preferences as their class is $\sim 2.25$ times better connected among each other than expected by chance, on the expense of weaker connectivity to remote classes. This effect is somewhat weaker for middle classes, which function as bridges between poor and rich classes, but apparently upward biased towards richer classes. This set of results directly verifies our earlier conjectures that the structure of the socioeconomic network is strongly stratified and builds up from social ties, whose creation is potentially driven by status homophily, and determined by the socioeconomic characteristics of individuals.

However, one can argue that the observed stratified structure can be simply the consequence of simultaneously present degree-degree and degree-wealth correlations. More precisely, if the degree of an individual is highly correlated with its economic status and at the same time the network is strongly assortative (i.e. people prefer to connect to other people with similar degrees) we may observe similar effects as in Fig.\ref{fig:3}a-c. To close out this possibility we completed an extensive correlation analysis, which showed us that no strong effects of degree-degree correlations can be detected and that the degree and wealth of individuals are very weakly correlated. To further clarify the effects of these correlations we performed a null model study where we carefully define random reference models to remove the correlations in focus in a controlled way and check their effects on the quantitative observations. As a conclusion we demonstrated that these correlations cannot explain the observed stratified structure. All of these results are presented in the Supplementary Materials (SM).

The above observations further suggest that the social structure may show assortative correlations in terms of socioeconomic status on the individual level. In other words, richer people may be better connected among themselves than one would expect them by chance and this way they form tightly connected ``rich clubs'' in the structure similar to the suggestion of  Mills \cite{Mills1956The}. This can be actually verified by measuring the rich-club coefficient \cite{Zhou2004Rich,Colizza2006Detecting}, after we adjust its definition to our system as follows. We take the original social network structure, sort individuals by their AMP value $P_u$ and remove them in an increasing order from the network (together with their connected links). At the same time we keep track of the density of the remaining network defined as
\begin{equation}
\phi(P_>)=\frac{2L_{P_>}}{N_{P_>}(N_{P_>}-1)}
\label{eq:phiP}
\end{equation}
where $L_{P_>}$ and $N_{P_>}$ are the number of links and nodes remaining in the network after removing nodes with $P_u$ smaller than a given value $P_>$. In our case, we consider $P_>$ as a cumulative quantity going from $0$ to $\sum_u P_u$ with values determined just like in case of $C_P(f)$ in Fig.\ref{fig:2}a but now using $100$ segments. At the same time, we randomise the structure using a configuration network model and by removing nodes in the same order we calculate an equivalent measure $\phi_{rn}(P_{>})$ as defined in Eq.\ref{eq:phiP} but in the uncorrelated structure. For each randomisation process, we used the same parameters as earlier and calculated the average density $\langle {\phi}_{rn}\rangle (P_{>})$ of the networks over $100$ independent realisations. Using the two density functions we define the "rich-club" coefficient as
\begin{equation}
\rho(P_>)=\frac{\phi(P_>)}{\langle {\phi}_{rn}\rangle(P_>)},
\label{eq:RCC}
\end{equation}
which indicates how many times the remaining network of richer people is denser connected than expected from the reference model. In our case (see Fig.\ref{fig:3}d purple symbols) the rich-club coefficient increases monotonously with $P_>$ and grows rapidly once only the richer people remain in the network. At its maximum it shows that the richest people are $\sim 8$ times more connected in the original structure than in the uncorrelated case. This provides a direct evidence about the existence of tightly connected ``rich clubs'' \cite{Mills1956The}, and the presence of strong assortative correlations in the social structure on the level of individuals in terms of their socioeconomic status. Note that this measure also suggests that the observed ``rich-clubs'' were not induced by degree-wealth correlations. The connectedness of nodes in the randomised structure were actually determined merely by their degrees, and since we kept wealth-degree correlations, the wealth-sorted removal process shows exactly the expected density of remaining richer nodes assuming only their original degree but no other correlations. This way the fraction of the two network density curves, i.e. the rich club coefficient, actually characterises exactly the effect of status homophily as compared to the randomised case where only degrees and degree-wealth correlations determined the connectedness of the network.

In addition, to rule out the possibility that our observation was induced by positive degree-degree correlations, we performed another randomisation of the network, where we kept node degrees, degree-degree, and degree-wealth correlations but vanished any other structural correlations. This randomisation procedure is a modification of the configuration network model and its definition is given in the SM. To measure the corresponding rich club coefficient function, we substituted in the numerator of Eq.\ref{eq:RCC} the residual network density function measured in this new degree correlated null model using the same wealth sorted removal sequence as earlier. Resutls in Fig.\ref{fig:3}d (black symbols) shows that the obtained rich-club coefficient appears approximately as a constant function around one. This way it demonstrates that the entangled effects of degree-degree and degree-wealth correlations cannot explain the emergence of ``rich-clubs'' observed in the empirical case. The network, which conserves degrees and these two correlations, emerges with a structure just as the network, which conserves degrees and degree-wealth correlations only. Consequently the observed increasing rich-club coefficient in case of the empirical structure is induced by status homophily or other tie creation mechanisms and not by degree-degree or degree-wealth correlations.

\subsection*{Spatial correlations between socioeconomic classes}

\begin{figure*}
\centering
\includegraphics[width=0.94\textwidth,angle=0]{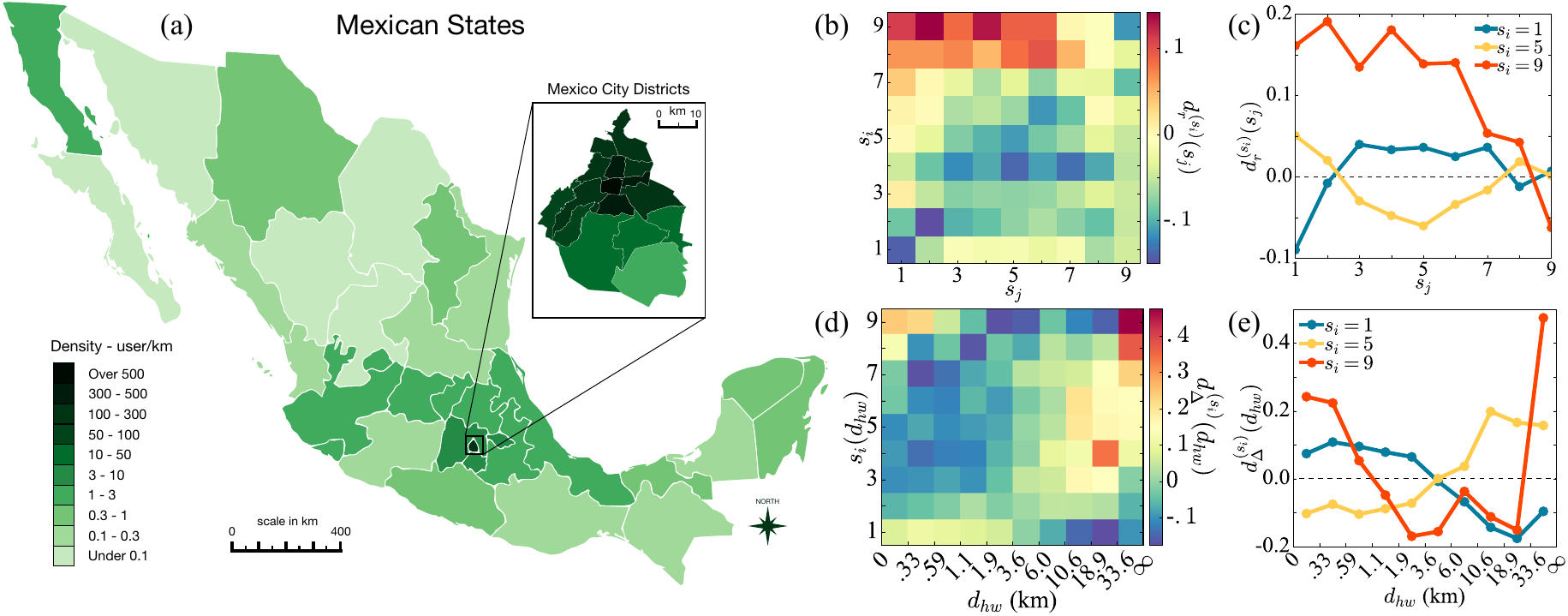}
\caption{Spatial socioeconomic correlations (a) State level population distribution of egos based on their zip locations. Inset depicts a zoom on the capital district. Informations depicted here were entirely obtained from the utilised dataset. The map representation was generated by using an open source code available at \href{https://gist.github.com/diegovalle/5843688}{github.com/diegovalle} (no copyright reserved) and shape files openly available at \href{http://www.inegi.org.mx}{www.inegi.org.mx} (no copyright reserved). (b) Relative average geodesic distances for different classes using the measure $d^{s_i}_{r}(s_j)$ defined in Eq.\ref{eq:dsi}. (c) The same $d^{s_i}_{r}(s_j)$ functions as on panel (b) shown for a selected set of classes (1-poor (blue), 5-middle (yellow), 9-rich (red)). (d) $d_{\Delta}^{s_i}(d_{hw})$ differences between commuting distance distributions calculated for different classes and for the whole population. $x$ scale depicts in logarithmic values of $d_{hw}$ commuting distances. (e) The same $d_{\Delta}^{s_i}(d_{hw})$ functions as on panel (d) shown for a selected set of classes (1-poor (blue), 5-middle (yellow), 9-rich (red)).
\label{fig:4}}
\end{figure*}

As we discussed earlier, the economic capacity of an ego strongly determines the possible places he/she can afford to live, arguably leading to somewhat homogeneous neighbourhoods, districts, towns, and regions occupied by people from similar socioeconomic classes. This effect may translate to correlations in the spatial distribution of socioeconomic classes in relation with each other.  To study such correlations, we use three different types of geographical informations extracted for individuals from the data: the zip code of reported billing address; the home; and work locations estimated from call activity logs (for details see Data and Materials). To give an overall image about the spatial distribution of the investigated users we use their zip location and assign them in different states of the country as shown in Fig.\ref{fig:4}a. Importantly, the observed population distribution correlates well with census data \cite{INEG2015} with coefficient $r=0.861$ ($p<.001$) on the state level, which indicates that our data records a fairly unbiased sample of the population in terms of distribution in space.

To quantify spatio-socioeconomic correlations, we measure the relative average geodesic distance between classes. More precisely, we take all connected egos $(u,v)\in E$ belonging to classes $u\in s_i$ and $v\in s_j$ respectively and measure the geodesic distance $d_{geo}^{zip}(a,b)$ between their zip locations. Using these values we calculate the average geodesic distance between any pairs of socioeconomic classes as
\begin{equation}
\langle d_{geo}(s_i,s_j)\rangle=\frac{1}{L(s_i,s_j)}\sum_{\substack{(u,v)\in E \\ u\in s_i, v \in s_j}}
d_{geo}^{zip}(u,v)
\label{eq:dgeo}
\end{equation}
where $L(s_i,s_j)$ assigns the number of links between nodes in classes $s_i$ and $s_j$. Note that since the social network is undirected the measure defined in Eq.\ref{eq:dgeo} is symmetric, i.e., $\langle d_{geo}(s_i,s_j)\rangle=\langle d_{geo}(s_j,s_i)\rangle$. Subsequently we calculate the average distance between nodes from class $s_i$ and any of their neighbours $\langle d_{geo}(s_i)\rangle$ to derive
\begin{equation}
d^{s_i}_{r}(s_j)=\frac{\langle d_{geo}(s_i,s_j)\rangle-\langle d_{geo}(s_i)\rangle}{\langle d_{geo}(s_i)\rangle}.
\label{eq:dsi}
\end{equation}
This measure is not symmetric anymore and gives us the relative average geodesic distance between egos in $s_i$ to egos in other classes $s_j$ as compared to the average distance of egos $s_i$ from any of their connected peers. Results are presented as a heat map matrix in Fig.\ref{fig:4}b where the diagonal component suggests a peculiar correlation. It shows that the relative average distance is always minimal (and negative) between egos of the same class $s_i$. This means that people tend to live relatively the closest to similar others from their own socioeconomic class as to egos from different classes, independently in which class they belong to. This is even more visible in Fig.\ref{fig:4}c after extracting the $d^{s_i}_{r}(s_j)$ curves (corresponding to rows in Fig.\ref{fig:4}b) for three selected classes. It highlights that while people of the poorest class live relatively the closest to each other, rich people tend to leave relatively the furthest from anyone from lower socioeconomic classes. These correlations are very similar to ones we already observed in the social structure suggesting that the stratified structure and spatial segregation may have similar roots. They are determined by the entangled effects of economic status and status homophily, together with other factors such as ethnicity or other environmental effects, which we cannot consider here.

Socioeconomic status of people may also correlate with their typical commuting distances (between home and work), a question which has been studied thoroughly during the last few decades. Some of these studies suggest a positive correlation between economical status (income) and the distance people travel every day between their home and work locations~\cite{Wheeler1967Occupational, Wheeler1969Some, Poston1972Socioeconomic}. Such correlations were partially explained by the positive payoff between commuting farther for better jobs, while keeping better housing conditions. On the other hand recent studies suggest that such trends may change nowadays as in central metropolitan areas, where the better job opportunities are concentrated, became more expensive to live and thus occupied by people from richer classes~\cite{LeRoy1983Paradise, Rosenthal2015Change}. Without going into details we looked for overall signs of such correlations by using the estimated home ($\ell_h$) and work ($\ell_w$) locations of individuals from different classes. For each ego we measure a commuting distances as $d_{hw}=|\ell_h-\ell_w|$ and compute the $P_{s_i}(d_{hw})$ distributions for everyone in a given $s_i$ class, together with the $P_{all}(d_{hw})$ distribution considering all individuals. For each class we are interested in
\begin{equation}
d_{\Delta}^{s_i}(d_{hw})=P_{s_i}(d_{hw})-P_{all}(d_{hw}),
\end{equation}
i.e., the difference of the corresponding distributions at each distance $d_{hw}$. This measure is positive (resp. negative) if more (resp. less) people commute at a distance $d_{hw}$ as compared to the overall distribution, thus indicating whether people of a given class are over (under)represented at a given distance. Interestingly, our data is in agreement with both above mentioned hypotheses, as seen in Fig.\ref{fig:4}d where we show $d_{\Delta}^{s_i}(d_{hw})$ for each class as a heat map. There, poorer people are over represented in shorter distances while this trend is shifted towards larger distances (see right skewed yellow component in Fig.\ref{fig:4}d) as going up in the class hierarchy. This continues until we reach the richest classes ($8$ and $9$) where the distance function becomes bimodal assigning that more people of these classes tend to live very far or very close to their work places as compared to expectations considering the whole population. This is even more visible in Fig.\ref{fig:4}e where selected $d_{\Delta}^{s_i}(d_{hw})$ functions are depicted for selected classes.

\section*{Discussion}

In this paper, we have investigated socioeconomic correlations through the analysis of a coupled dataset of mobile phone communication records and bank transaction history for millions of individuals over $8$ months. After mapping the social structure and estimating individual economic capacities, we addressed four different aspects of their correlations: (a) we showed that individual economic indicators such as average monthly purchases and also debts are unevenly distributed in the population in agreement with the Pareto principle; (b) after grouping people into nine socioeconomic classes we detected effects of status homophily and showed that the socioeconomic network is stratified as people most frequently maintain social ties with people from their own or neighbouring social classes; (c) we observed that the social structure is upward-biased towards wealthier classes and show that assortative correlations give rise to strongly connected ``rich clubs'' in the network; (d) finally, we demonstrated that people of the same socioeconomic class tend to live closer to each other as compared to people from other classes, and found a positive correlation between their economic capacities and the typical distance they use to commute.

Even though our study is built on large and detailed data, the utilised data covers only partially the population of the investigated country. However, as we demonstrated above, for population-level measures, such as the Gini coefficient and spatial distribution, we obtained values close to independently reported cases, and thus our observations may generalise in this sense. In addition, the question remains how well mobile phone call networks approximate the real social structure. A recent study \cite{Eagle2009Inferring} demonstrated that real social ties can be effectively mapped from mobile call interactions with precision up to $95\%$. However, it is important to keep in mind that the poorest social class of the society is probably under-represented in the data as they may have no access to bank services and/or do not hold mobile phones. Datasets simultaneously disclosing the social structure and the socioeconomic indicators of a large number of individuals are still very rare. However, several promising directions have been proposed lately to estimate socioeconomic status from communication behaviour on regional level \cite{Specanovic2015Mobile, Blumenstock2010Mobile, Mao2015Quantifying} or even for individuals \cite{Blumenstock2015Predicting}, just to mention a few. In future works these methods could be used to generalise our results to other countries using mobile communication datasets. Here, our aim was to report some general observations in this direction using directly estimated individual economic indicators. Our overall motivation was to empirically verify some long-standing hypothesises and to explore a common ground between hypothesis-driven and data-driven research addressing social phenomena.

\section*{Data and Materials}

\subsection*{Mobile communication data}
Communication data used in our study records the temporal sequence of 7,945,240,548 call and SMS interactions of 111,719,360 anonymised mobile phone users for $21$ months (between January 2014 and September 2015) in Mexico. Each call detailed record (CDR) contains the time, unique caller and callee IDs, the direction and duration of the interaction, and the cell tower location of the client(s) involved in the interaction. Other mobile phone users, who are not clients of the actual provider also appear in the dataset with unique IDs. All unique IDs are anonymised as explained below, thus individual identification of any person is impossible from the data. Using this dataset we constructed a large social network where nodes were users (whether clients or not of the actual provider), while links were drawn between them if they interacted (via call or SMS) at least once during the observation period. In order to filter out call services and other non-human actors from the social network, after construction we recursively removed all nodes (and connected links) who appeared with either in-degree $k_{in}=0$ or out-degree $k_{out}=0$. We repeated this procedure recursively until we received a network where each user had $k_{in}, k_{out}>0$, i.e. made at least one outgoing and received at least one incoming communication events during the nearly two years of observation. After construction and filtering the network remained with 82,453,814 users connected by 1,002,833,289 links, which were considered to be undirected after this point.

\subsection*{Credit and purchase data}

To estimate individual economic indicators we used a dataset provided by a single bank in the studied country. This data records financial details of $6,002,192$ of people assigned with unique anonymised identifiers over $8$ months from November 2014 to June 2015. The data provides time varying customer variables as the amount and type of their daily debit/credit card purchases, their monthly loan measures, and static user attributes as their billing postal code (zip code), their age, and gender. In addition for a subset of clients we have the records of monthly salary (38.9\% of users) and income (62.5\% of users) defined as the sum of their salaries and any incoming bank transactions. Note that the observation period of the bank credit informations falls within the observation period of the mobile communication dataset, this way ensuring the largest possible overlap between the sets of bank and mobile phone customers.

\subsection*{Location data}
We used two types of location data for a set of customers. We used the zip code of billing address of bank customers (also called zip location). We also estimated the work and home locations for a set of users using geo-localised mobile communication events. To determine home (resp. work) locations we looked for the most frequented locations during nights and weekends (resp. during daylight at working days). From the total 992,538 individuals we found 990,173 with correct zip codes, and 94,355 with detectable home and work locations (with at least 10 appearances at each location). Each method has some advantages and disadvantages. While frequency dependent locations are more precise, they strongly depend on the activity and regularity of users in terms of mobility. On the other hand, zip codes provide a more coarse-grained information about the location of individuals but they are assumed to be more reliable due to reporting constraints to the bank and because they do not depend on the call activity of individuals.

\subsection*{Combined datasets and security policies}

A subset of IDs of the anonymised bank and mobile phone customers were matched. The matching, data hashing, and anonymisation procedure was carried out through direct communication between the two providers (bank and mobile provider) and was approved by the national banking commission of the country. This procedure was done without the involvement of the scientific partners. After this procedure only anonymised hashed IDs were shared disallowing the direct identification of individuals in any of the datasets. Due to the signed non-disclosure agreements and the sensitive nature of the datasets it is impossible to share them publicly.

This way of combining of the datasets allowed us to simultaneously observe the social structure and estimated economic status of the connected individuals. The combined dataset contained 999,456 IDs, which appeared in both corpuses. However, for the purpose of our study we considered only the largest connected component of this graph containing IDs valid in both data corpuses. This way we operate with a connected social graph of 992,538 people connected by 1,960,242 links, for all of them with communication events and detailed bank records available.

\section*{Competing financial interests}
We have no competing interests.

\section*{Author contributions}
All authors participated in the design of the project and the writing of the manuscript. Y.L., M.K. and E.F. designed the measures, Y.L., and M.K. performed the data analysis. All authors reviewed the manuscript.

\section*{Acknowledgements}
We thank for M. Fixman and J. Brea for assistance with the data set and for J. Saram\"aki and J.P. Chevrot for useful discussions and for the anonymous reviewers for their constructive comments.

\section*{Funding}
We acknowledge the support from the SticAmSud UCOOL project, INRIA, and the CODDDE (ANR-13-CORD-0017-01) and SoSweet (ANR-15-CE38-0011-01) ANR projects.

\pagebreak

{\LARGE \textbf{Supplementary Materials}}

\addtocontents{toc}{\protect\setstretch{0.1}}

\section{Degree and wealth correlations}

In the main text we studied a social network where each individual was assigned with a socioeconomic indicator defined as their average monthly purchase (AMP) (see Eq.1 in the main text).  We used these indicators to estimate the socioeconomic status of individuals and group them into $9$ exclusive socioeconomic classes. By analysing the social network and the assigned socioeconomic classes we observed that individuals tend to connect to similar others from their own or neighbouring socioeconomic classes, while social ties with people from remote classes are less frequent. We argued that this observed stratification in the social structure (see in Fig.3 a-c in the main text) is due to the entangled effects of socioeconomic imbalances and status homophily, i.e. the tendency of people to connect to others with similar socioeconomic status.

However, one can argue that the observed stratified structure can be simply the consequence of simultaneously present degree-degree and degree-wealth correlations. More precisely, if the degree of an individual is highly correlated with its economic status (wealth) and the network is strongly assortative (i.e. people prefer to connect to other people with similar degrees) we may observe similar effects as in Fig.1 (in the main text). To close out this possibility we present here a correlation analysis and a null model study where we carefully define random reference models to remove the correlations in focus in a controlled way and check their effects on the quantitative observations.

\subsection{Degree-degree correlations}

\begin{figure}[ht!] \centering
  \includegraphics[width=.68\textwidth]{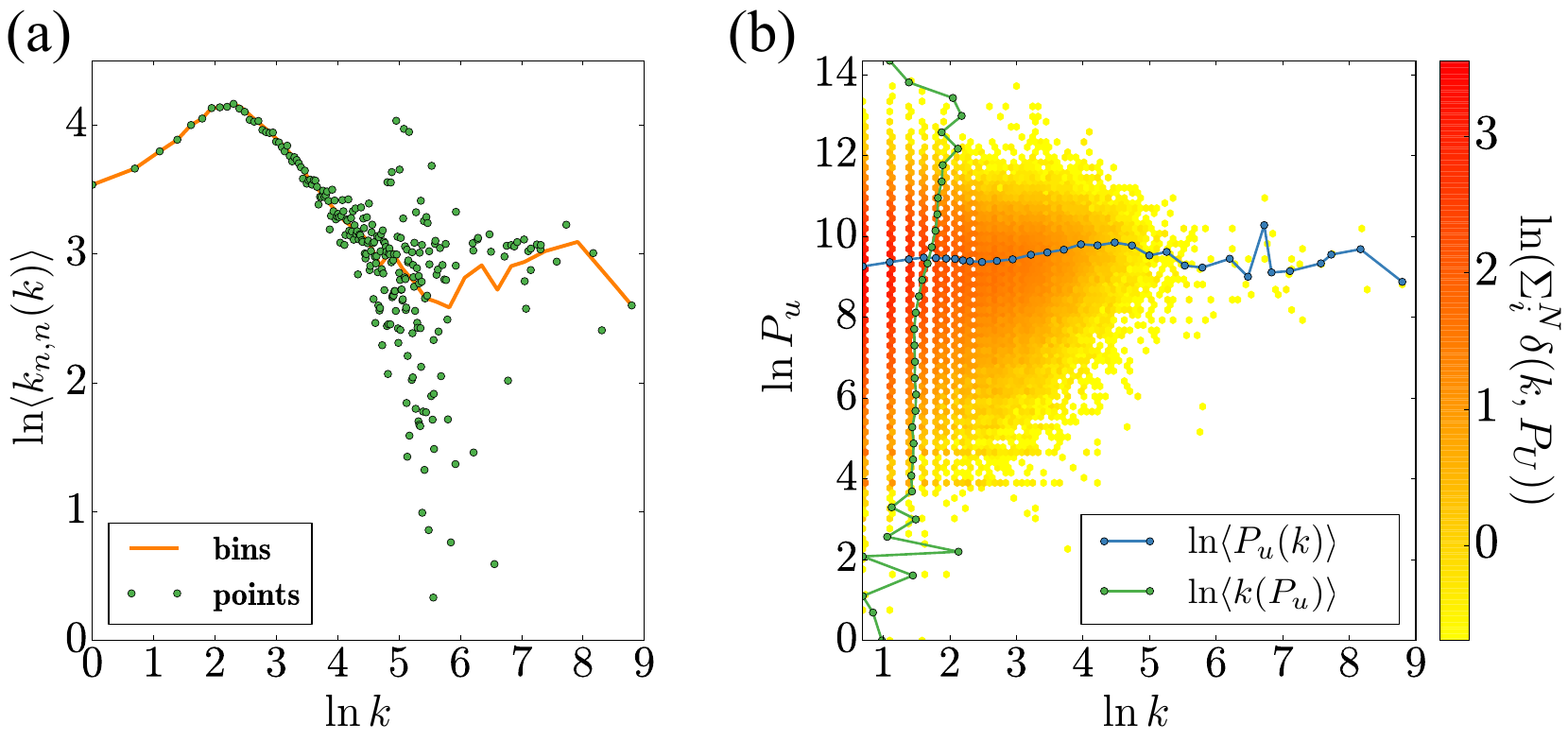}
\caption{\small Degree-degree and degree-wealth correlations in the social-economical network. (a) The $k_{nn}(k)$ function computed in the social network. (b) Correlation plot (shown as a heat map) of degree-wealth correlations. Blue (resp. green) horizontal (resp. vertical) solid line and symbols show the average wealth (resp. degree) as the function of degree (resp. wealth).}
\label{fig:S1}
\end{figure}

The simplest way to characterise degree-degree correlations in a network is by computing the Pearson correlation coefficient between two random variables identified as the degrees of nodes connected in the network structure \cite{Newman2010Networks}. After calculating the Pearson correlation coefficient in the investigated social network we found that it is $r\approx -0.00813 $ ($p<0.001$, $SE=7.13\times 10^{-4}$), suggesting that the network shows no (or very weak disassortative) degree-degree correlations. However, since the Pearson correlation coefficient gives only an overall characteristic measure and assumes that correlations are linear, we further investigate degree-degree correlations with another metric, which is conventionally used to characterise degree-degree correlations. We measure the $k_{nn}(k)$ average nearest neighbour degree for each degree class $k$ in the network \cite{Newman2010Networks}. This function (shown in Fig.S\ref{fig:S1}a) disclosed a somewhat more sophisticated picture about degree-degree correlations. First of all it shows that it is not a monotonous function but it assigns mixed effects of assortative and disassortative mixing. It shows positive (assortative) correlations up to $k\simeq 10$, which after it indicates negative (disassortative) correlations, and becomes flat for the largest degrees. Consequently our network does not show strong assortative correlations over its whole degree range, which suggest that degree correlations may not evidently play a deterministic role in the observed stratified structure even if they are correlated with wealth. Note that this type of complex functional scaling of $k_{nn}(k)$ commonly characterises non-mutualised directed networks as discussed in \cite{Li2014A}. This is in line with our case where the network was not mutualised were kept in the structure, which were assumed to be indirected after the network construction for the convenience of our study.

\subsection{Degree-wealth correlations}

Entangled with degree correlations, dependencies between node degree and economic status (wealth) can also contribute to the emergence of the observed stratified structure. To characterise this correlation, first we measure again the Pearson correlation coefficient between the degree $k$ and AMP value $P_u$ of each node. This correlation turns out to be small with coefficient $r\approx 0.0357$ ($p<0.001$, $SE=9.71\times 10^{-1}$). To obtain a more complete picture about their dependencies we simply show in Fig.\ref{fig:S1}b the binned scatter plot as a heat map of these two variables and in addition calculate the average value of wealth (resp. degree) as the function of degree (resp. wealth). These results indicates the weak dependencies between these variables for the whole range of variables and although un-disclose some non-monotonous dependencies between node degree and economic status but do not indicate evident strong positive correlations between these variables.

\subsection{Null model study}

\begin{figure}[ht!] \centering
  \includegraphics[width=.68\textwidth]{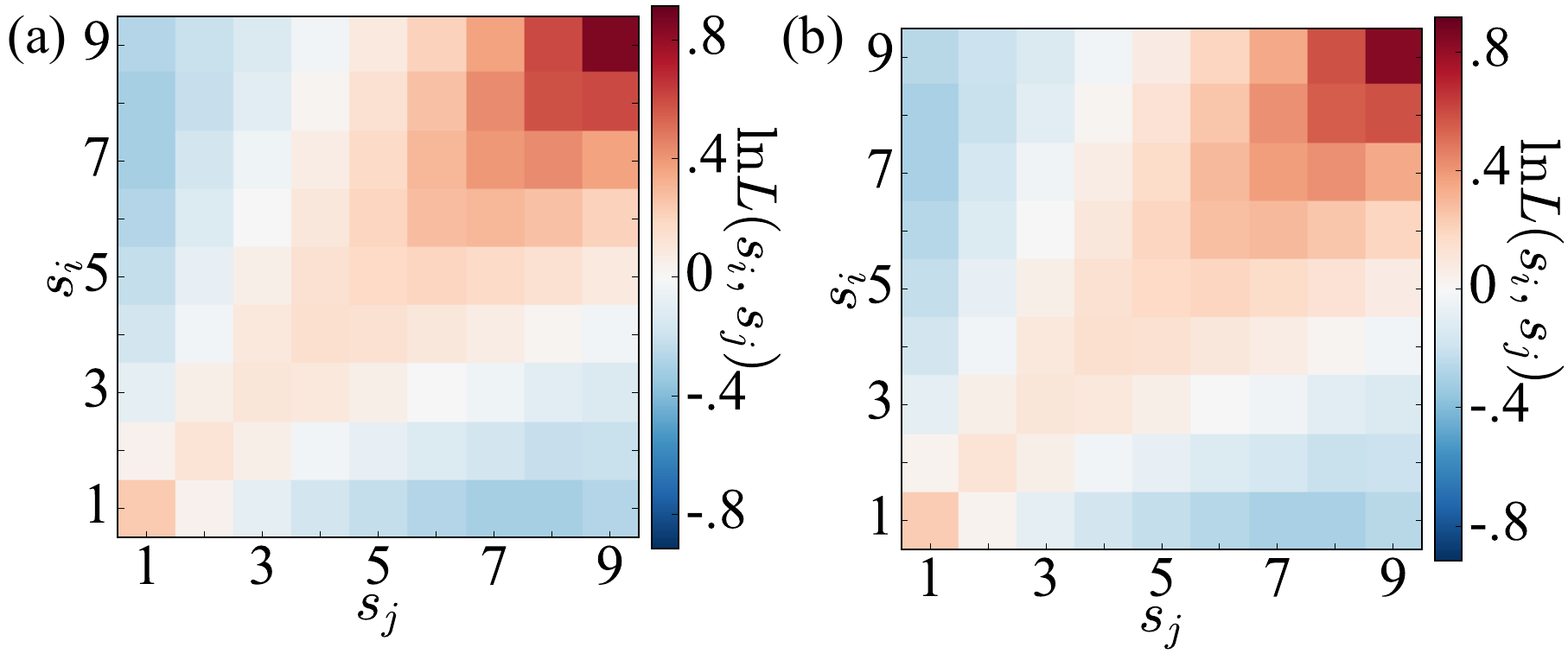}
\caption{\small $L(s_i,s_j)$ normalised socioeconomic class connectivity matrices in case of two null models (for definition see Eq.4 in the main text). In each case the numerator was taken as the socioeconomic class connectivity matrices of the original network, while the denumerator was measured from a null model structure of (a) NM1, (b) NM2.}
\label{fig:S2}
\end{figure}

We concluded that the social structure is stratified by socioeconomic status by measuring the fraction $L(s_i,s_j)=|E(s_i,s_j)|/|E_{rn}(s_i,s_j)|$ (see also in Eq.4 in the main text) of the number of links connecting people from different classes $s_i$ and $s_j$ in the original structure and in a null model structure. In this case the null model structure is defined as the configuration network model of the empirical network (here we call null model 1 (NM1)), where we take the original social network, select random pairs of links and swap them without allowing multiple links and self loops. In order to remove any residual correlations we repeated this procedure $5\times |E|$ times (where $|E|$ being the number of links in the social network). This randomisation keeps the number of links, individual economic indicators $P_u$, and the assigned class of people unchanged, but destroys degree-degree correlations, possibly present community structure (for a summary of present correlations see Table S\ref{table:S1}). Since it destroys all possible structural correlations in the social network (apart from correlations due to unavoidable finite size effects) as a consequence it eliminates the socioeconomic layers as well. In each case, we repeat this procedure for $100$ times and present results averaged over the independent random realisations. Results shown in Fig.S\ref{fig:S2}b appears with a diagonal component, which evidently assigns strong connectivity between neighbouring socioeconomic classes, i.e. a stratified structure. This measure simply assigns \textit{how many times people of different classes are more connected as compared to the case where they are connected by chance}. The observed diagonal component indeed assigns the significance of this correlations what we associated to status homophily.

To further investigate potential reasons behind this observation, let's take a null model to directly address the possible effects of degree homophily. In this null model (NM2) we destroy the potential community structure and all other structural correlations including socioeconomic layers, but conserve degree-degree and degree-wealth correlations (see Table S\ref{table:S1}). NM2 is defined as a modification of the configuration network model, where instead of selecting link pairs randomly to swap, we select a link and one of its end randomly, and choose another link randomly where the degree of one of the ending nodes is equal to the degree of the selected end of the first link. Swapping the other ends of the links (with potentially different degrees) will result yet two links between nodes of the original degrees but connected randomly otherwise. We do not allow self loops and multiple links and skip to swap links where both node degrees are unique in the network. We found $22$ such cases from $\sim 2M$ links thus we assume this condition will not bias our shuffling considerably. We swapped randomly link pairs $5\times |E|$ just as for the NM1 model and computed averages over $100$ realisations.

\begin{table}[h]
\centering
  \begin{tabular}{| c ||c | c | c |}
\hline
 & Original & NM1 & NM2 \\ \hline \hline
degree-degree & \checkmark & $\times$ & \checkmark \\ \hline
degree-wealth & \checkmark & \checkmark & \checkmark \\ \hline
communities & \checkmark & $\times$ & $\times$ \\ \hline
  \end{tabular}
\caption{\small Correlations present in different null models. We consider degree-degree, degree-wealth, and higher-order structural (communities) correlations in the original network (Orig) and two null models (for definitions see text).}
\label{table:S1}
\end{table}

Our hypothesis is that if the present degree-degree correlations in NM2 would explain the observed stratified structure, then after using the corresponding $|E_{rn}^{NM2}(s_i,s_j)|$ link density matrix in the normalisation of $L(s_i,s_j)$ (see Eq.4 in the main text) the resulting matrix should become flat. This would mean that the actually present correlations could explain (reproduce) the empirical observations. However, this is not the case here as seen in Fig.S\ref{fig:S2}b. The $L(s_i,s_j)$ matrix normalised by the corresponding NM2 matrix appears to be almost identical than the one normalised by the NM1 matrix. This suggests that degree-degree correlations and degree homophily do not play a role here, thus it cannot explain the emergence of the stratified structure.

\end{document}